\documentclass[pre,twocolumn,aps,superscriptaddress,showpacs,showkeys]{revtex4-1}
\usepackage{amsmath,amssymb,amsfonts,bm,amsthm}
\usepackage{graphicx}
\usepackage{times}
\usepackage{float}
\usepackage{lineno}
\usepackage{color}
\usepackage{amsthm}
\usepackage{soul}
\usepackage[english]{babel} 

\usepackage{tabularx} 

\theoremstyle{definition}

\usepackage[colorlinks=true, linkcolor=blue, citecolor=blue, filecolor=blue, runcolor=blue, urlcolor = blue]{hyperref}

\usepackage{graphicx}

\begin{document}

\title{Topological Methods for Polymeric Materials: Characterizing the Relationship Between \\ Polymer Entanglement and Viscoelasticity.}

\author{E. Panagiotou}
\email[Corresponding author: ]{lnpanagiotou@yahoo.com}
\affiliation{Department of Mathematics and SimCenter, University of Tennessee at Chattanooga, TN 37403}

\author{K. C. Millett}
\affiliation{Department of Mathematics, University of California Santa Barbara, CA 93106-3080}

\author{P. J. Atzberger}
\email{atzberg@gmail.com}
\homepage[Website: ]{http://atzberger.org/}
\affiliation{Department of Mathematics and Department of Mechanical Engineering, University of California Santa Barbara, CA 93106-3080}

\begin{abstract}
We develop topological methods for characterizing the relationship between polymer chain entanglement and bulk viscoelastic responses.  We introduce generalized Linking Number and Writhe characteristics that are applicable to open linear chains.  We investigate the rheology of polymeric chains entangled into weaves with varying topologies and levels of chain density.  To investigate viscoelastic responses, we perform non-equilibrium molecular simulations over a range of frequencies using sheared Lees-Edwards boundary conditions.  We show how our topological characteristics can be used to capture key features of the polymer entanglements related to the viscoelastic responses.  We find there is a linear relation over a significant range of frequencies between the mean absolute Writhe $Wr$ and the Loss Tangent $\tan(\delta)$.  We also find an approximate inverse linear relationship between the mean absolute Periodic Linking Number $LK_P$ and the Loss Tangent $\tan(\delta)$.  Our results show some of the ways topological methods can be used to characterize chain entanglements to better understand the origins of mechanical responses in polymeric materials.
\end{abstract}


\maketitle

\section{Introduction}
\label{sec:introduction}

A central aim in polymeric material science is to understand the relationships between chemistry, molecular-level interactions, and bulk material properties.  We consider polymeric materials and develop approaches for investigating the relationship between the topology of polymeric chain interactions and the resulting bulk viscoelastic mechanical responses.  The collective configurations of the chains can result in complex entangled structures that greatly restrict chain motions and augment transmission of mechanical stresses within the material~\cite{Doi1986,deGennes1979,Rubinstein2003}.  Characterizing the intuitive notion of entanglement in a way that is quantitatively precise poses many interesting challenges.  For instance, entanglement in polymer melts or gels often involves kinetics over a broad range of time-scales with collective responses that often depend on both the frequency and duration of the mechanical perturbations.  In addition, the polymer chains often exhibit friction with respect to one another or coupling from immersion within a solvent fluid. This can result in significant variations in the viscoelastic responses that may depend sensitively on frequency~\cite{Bird1987Vol1,Bird1987Vol2,Doi1986}.

When investigating polymer chain entanglement, a common approach is to consider two length and time scales.  First, there is the length and time scale of the entire chain, where global entanglement occurs as the chains get knotted and linked to each other \cite{Doi1986,Edwards1967,Edwards1968,Kremer1990,Kim2008,Baig2010}. Second, one has local entanglement arising from the local constraints, obstacles, that a chain feels over a small length and time scale. This entanglement length often plays a central role in models of entangled polymer dynamics and is closely related to the tube diameter in tube model theories \cite{Hoy2009,Rubinstein2003,Shanbhag2007,PPprojection,Stephanou2010}. Understanding the relation between these scales poses a number of challenges in practice.  For instance, experimentally there are different methods for the determination of the plateau modulus and different entanglement molecular weights give different results \cite{Liu2006}. Since Edwards' tube model, several improvements of this theory have emerged, but inconsistencies still remain \cite{Likhtman2002,Everaers2011,Desai2016,Snijkers2015}. Moreover, an even shorter length scale, that of the packing length, has been shown to have an influence and has been incorporated in some theories of polymer viscoelasticity \cite{Unidad2015,Fetters1999,Sussman2012}.

The relation between entanglement and viscoelastic properties of materials has been studied indirectly by varying the density or molecular weight of the chains thereby influencing the number of contacts between neighboring chains ~\cite{Rubinstein1984, Jain2012,Everaers2004, Kroeger2005,Tzoumanekas2006}.
A measure of polymer entanglement that has been very helpful in such studies is the number of "kinks" per chain, which is derived from the application of a contour reduction algorithm on a polymer melt \cite{Kroeger2005,Tzoumanekas2006,Everaers2004,Bisbee2011}. However, the global entanglement complexity is more subtle and cannot be assessed by only measuring the number of contacts.  This has led to the use of tools from knot theory to study entanglement in polymers \cite{Edwards1967,Edwards1968,Foteinopoulou2008,Everaers2004,Qin2011,Kuei2015,Everaers1996}. The difficulty in using tools from knot theory is that they are defined on simple closed curves in space (rings) while the polymers often have other architectures, such as being open linear chains. Toward dealing with this issue, a new statistical definition of knotting was introduced in \cite{Millett2004}. This method can determine the principal knot type of a fixed configuration of an open chain.  However, the polymer chains move in time and this method may have problems when applied to study entanglement in non-equilibrium conditions.  Moreover, the tools from knot theory have not yet been developed sufficiently for polymeric systems with three-dimensional Periodic Boundary Conditions (PBC) which are often used in practice \cite{Morton2009}.

In our work, we take a complimentary approach building on mathematical ideas from topology and geometry to quantify the complexity of the polymer entanglements.  We introduce two quantities referred to as the collective ``Linking Number" and ``Writhe"~\cite{Panagiotou2015,Panagiotou2013b,Panagiotou2014}. The advantage of the Gauss linking integral is that it can be applied to both linear and ring polymers and it is a continuous function of the chain coordinates. Moreover, in \cite{Panagiotou2015} it has been shown that it can be extended to systems with 1,2 or 3 PBC to provide a new continuous measure of entanglement. These measures have been applied to study polymer entanglement in both equilibrium and non-equilibrium conditions. More precisely, it was shown that the writhe in combination with the Z1 algorithm can provide a new estimator of the entanglement length with several advantages over other estimators and our proposed measures have also been used to understand the dis-entanglement of polymer chains in a melt under an elongational force \cite{Panagiotou2011,Panagiotou2013b,Panagiotou2014}.  Recently, the Gauss linking integral has been also used to study protein folding kinetics \cite{Panagiotou2018b}. These results indicate the promise of such topology-based estimators in polymer theories.

Here, we develop further these topological approaches for characterizing polymer entanglements within non-equilibrium systems subject to external materials deformations, such as oscillatory shearing.  To investigate viscoelastic responses, we develop methods for non-equilibrium three-dimensional molecular simulations with shearing Lees-Edwards periodic boundary conditions (LE-PBC)~\cite{Lees1972}.  We use our approach to study the frequency dependent viscoelastic responses and relationships to the underlying topology of the polymer chain entanglements.

We consider in our investigations polymeric weaves which have well-controlled topological properties that we can vary.  We consider polymeric systems that have local topologies arising from short linear polymer chains and those with global topologies arising from long linear polymer chains.  We also vary density to consider systems ranging from weakly entangled to strongly entangled.  Our topological approaches allow us to investigate both the local and global entanglement effects.  We find an approximate linear relation over a large range of frequencies between the mean absolute Writhe $Wr$ and the loss tangent $\tan(\delta)$.  We also find an approximate inverse linear relationship between the mean absolute Periodic Linking Number $LK_P$ and the loss tangent $\tan(\delta)$.  We expect our topological approaches could be useful in gaining additional quantitative information relevant to understanding the mechanics of polymeric materials.

We organize the paper as follows. In Section \ref{LK} we introduce approaches from knot theory to precisely characterize the topology of the polymeric chains.  In Section \ref{weavessys} we describe a class of polymeric materials having a weave-like topology.  In Section \ref{simmethod}, we discuss the details of our computational methods and simulation approaches.  In Section \ref{res}, we present our results and discuss relationships revealed by our methods between the topology of the polymer entanglements and bulk material responses.

\section{Characterizing Polymer Entanglement} \label{LK}

We measure the degree to which polymer chains interwind and attain complex configurations using the Gauss Linking Integral~\cite{Gauss1877}.  We define the Gauss \textit{Linking Number} as 
\begin{eqnarray}
\label{Gausslk} 
\label{lk}
\\
\nonumber
L(l_1,l_2)=\frac{1}{4\pi}\int_{[0,1]}\int_{[0,1]}\frac{(\dot\gamma_1(t),\dot\gamma_2(s),\gamma_1(t)-\gamma_2(s))}{||\gamma_1(t)-\gamma_2(s)||^3}dt ds.
\end{eqnarray}
We use this for two disjoint (closed or open) oriented curves $l_1$ and $l_2$ whose arc-length parametrizations are respectively $\gamma_1(t),\gamma_2(s)$. The double integral is over $l_1$ and $l_2$.  In this notation $(\dot\gamma_1(t),\dot\gamma_2(s),\gamma_1(t)-\gamma_2(s))$
denotes the \textit{scalar triple product} of $\dot\gamma_1(t),\dot\gamma_2(s)$ and
$\gamma_1(t)-\gamma_2(s)$.
The Gauss Linking Number is a topological invariant for closed chains and a continuous function of the chain coordinates for open chains.
We also define a one chain measure for the degree of intertwining of the chain around itself.  

We define the \textit{Writhe} of a chain as
\begin{eqnarray}
\label{Gausslk}
\label{Wr_def}
\\
\nonumber
Wr(l)=\frac{1}{4\pi}\int_{[0,1]}\int_{[0,1]}\frac{(\dot\gamma(t),\dot\gamma(s),\gamma(t)-\gamma(s))}{||\gamma(t)-\gamma(s)||^3}dt ds.
\end{eqnarray}
For a curve $\ell$ with arc-length parameterization $\gamma(t)$ is the double integral over $l$.  The Writhe is a continuous function of the chain coordinates for both open and closed chains.

For systems employing Periodic Boundary Conditions (PBC), the linking that is imposed from one simulated chain on another chain propagates in three dimensional space by the images of the other chain.  In other words, for a system with PBC each simulated chain gives rise to a \textit{free chain} in the periodic system which consists of an infinite number of copies of the simulated chain. We call each copy of a chain an \textit{image} of the free chain. It has been shown that a measure of entanglement that can capture the global linking in a periodic system is the \textit{periodic linking number} $LK_P$ \cite{Panagiotou2015}:

We define the \textit{Periodic Linking Number} as
\begin{eqnarray}
\label{lkpdef1}
\label{lk3}
LK_P(I,J)=\sum_{v}L(I_u,J_v),
\end{eqnarray}
The $I$ and $J$ denote two (closed, open or infinite) free chains in a periodic system. Suppose that $I_u$ is an image of the free chain $I$ in the periodic system. The sum is taken over all the images $J_v$ of the free chain $J$ in the periodic system.  We say that the \textit{Periodic Linking Number} $LK_P$ is between two free chains $I$ and $J$.

The Periodic Linking Number is a topological invariant for closed chains and a convergent series for open chains that changes continuously with the chain coordinates. For its computation, we use a cutoff, the \textit{local Periodic Linking Number}~\cite{Panagiotou2011,Panagiotou2015}.

\section{Polymeric Materials with Weave-like Entanglements}\label{weavessys}

\begin{figure}[H]
\centering
\includegraphics[width=0.9\columnwidth]{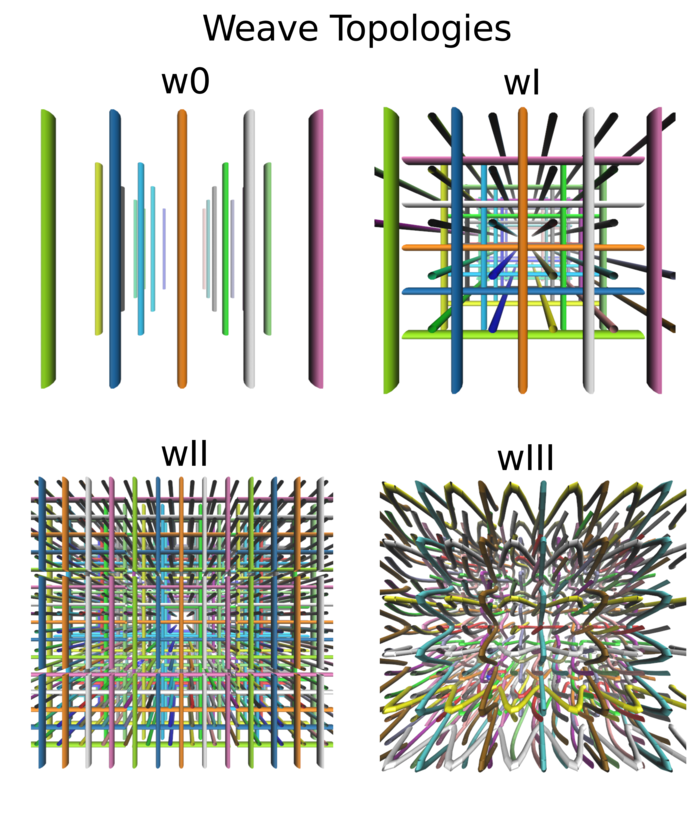}
\caption{We consider polymeric chains entangled with weave-like topologies.  The weave0 (w0) denotes case with aligned chains, weaveI (wI) the case with smaller density of orthogonal and non-interlaced chains, weaveII (wII) the case with larger density of orthogonal and non-interlaced chains, and weaveIII (wIII) the case with alternating interlaced chains.}
\label{fig:configex00}
\end{figure}

\begin{figure}[H]
\centering
\includegraphics[width=0.95\columnwidth]{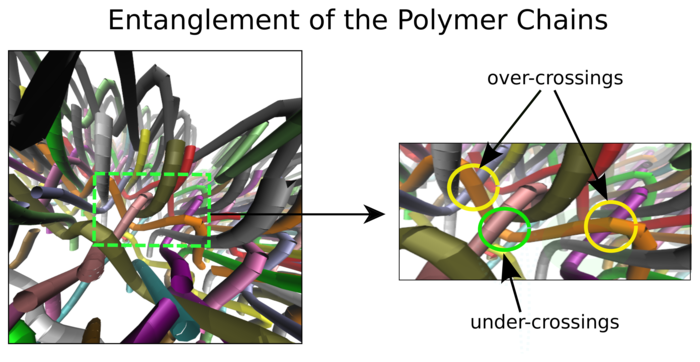}\\
\caption{We show the entaglements of the polymer chains of wIII. The weave wIII has a topology with alternating interlaced chains. We see the chains in the x direction which alternatingly go over and under chains in the perpendicular y direction.  We show one chain in the x-direction (orange curve) that can be seen locally to meet with three chains in the perpendicular y direction. }
\label{fig:configex001}
\end{figure}

We study the role of entanglement topology in the mechanical responses of polymeric materials.  We consider both the case when the global topology is fixed and when the topology can change over time for a few different weave entanglements, see Figures \ref{fig:configex00} and \ref{fig:configex001}.  We investigate how mechanical responses depend on the topology, chain density, or  whether the polymers are to be considered open chains or closed (infinite) chains.

We refer to chains as closed when they are very large relative to the length-scale of the entanglements.  In practice, we can think of these as effectively infinitely long chains.  These chains have the important property that in computations no end-points occur within the simulation box.  These chains are treated as extending periodically to create a toopology corresponding to an infinite weave.  

We refer to chains as open when they are finite in length.  These chains are open in the sense they always have end-points within the simulation box.  These chains can still cross the periodic boundary where they interact with the periodic image points generated by the unit cell.  In this case, the topology of the material simulated can change over time in response to the  deformations and stresses of the material.  

We consider in our studies the specific weave topologies referred to as (w0) for aligned, (wI, wII) for orthogonal non-interlaced at different densities and (wIII) alternating interlacing.  We show the base-line chain density and lengths in Table \ref{table_weavetypes}.  We show example configurations of each of these weaves in Figures \ref{fig:configex00}, \ref{fig:configex001}.

We start with the weave w0 which has a relatively simple global topology.  The polymers are simply arranged parallel to one another without any entanglement, providing a good reference case for topology and responses.  We then consider the weaves wI and wII that arrange polymers in a regular criss-cross patterm.  Both wI and wII have the same global topology, but we take these to have different densities.  We take weave wIII to be an inter-woven topology alternating in-out entanglements.  We take wIII to have the same density as wII.  The weave wIII has a non-trivial global topology.  As a result of the polymer sterics preventing the crossing of chains, the global topology of the infinite systems cannot change without breakage of the bonds.  It should be noted that the local configurations of the chain interactions
can lead to local entanglements that change over time under the global constraints of the topology.

We create open chain systems for w0, wI, wII, wIII by deleting the same bond from each chain in the simulation box.  Given the periodicity this creates a standard procedure for obtaining an initial open-chain configuration. As a result of the polymer sterics preventing the crossing of chains, the global topology of the infinite systems cannot change without breakage of the bonds.  In contrast to the infinite chain systems, the global topology of the open chain system can change over time by slippage of the chains past the entanglements.  This creates the possibility in response to mechanical reformations for topology rearrangements over time.  

We investigate the mechanical responses of the material by subjecting the polymeric chains to an oscillatory shear through deformation of the simulation box in the style of Lees-Edwards~\cite{Lees1972}.  This provides us with shear stresses for the material which we can correlate with simultaneous measurements of the chain density and topology of the materials.  We remark that similat to our polymeric weave configurations, there have also been studies using weaves for investigating metal organic frameworks and crystals~\cite{Evans2013}.

\begin{table}[H]
\centering
\begin{tabular}{|l|l|l|l|}
\hline
\textbf{Weave}           & \textbf{Topology}                & \textbf{Density}         & \textbf{MW (open)}\\
\hline
W0             & parallel, non-interlaced         & 0.0625 (15 $amu/nm^3$)   &  20 $m_0$\\
WI             & orthogonal (non interl.)      & 0.1875 (45 $amu/nm^3$)   &  20 $m_0$\\
WII           & orthogonal (non interl.)      & 0.33 (80 $amu/nm^3$)     &  15 $m_0$\\
WIII         & alternating interlaced             & 0.35 (84 $amu/nm^3$)     &  21-17 $m_0$\\
\hline
\end{tabular}
\caption{Densities associated with the polymeric weaves shown in Figure \ref{fig:configex00}}
\label{table_weavetypes}
\end{table}

\section{Simulation of the Polymers}\label{simmethod}

We investigate entangled polymeric chains using three-dimensional molecular simulations.  The polymers are treated as elastic macromolecules modeled with harmonic bond potential of energy $E=K_b(r-r_0)^2$, $K_b=250$,  $r$ denotes the length of extension of the polymer bonds and $r_0=1$ denotes the rest length of the bond. The polymer bending stiffness is controlled with a harmonic angle potential with energy $E=K_{\theta}(1-\cos(\theta-\theta_0))$, with $K_{\theta}=8$, where $\theta$ is the angle between two consecutive bonds. The rest angle is $\theta_0=\pi$. The length of the simulation box is approximately $20\sigma$.  Each polymer chain has approximately $20$ monomers inside the simulation box for the densities and parameters given in Tables \ref{table_weavetypes} and \ref{tableparams1}.  With this choice of $K_{\theta}$, the chains have persistence length of approximately $1/5$ of the length of the simulation box. With these potentials, there is no maximum permitted length or bond angle constraints, but there is a high energy penalty for large deviations from the rest length.  This does not exclude the possibility of chains crossing through each other, especially for large deformations. Our results however, show that chain crossings are rare enough so as to not influence the qualitative effects of entanglement observed here (see Section \ref{res}).  The beads of our polymers interact through the Lennard-Jones (LJ) potential with energy
\begin{equation}
E_{LJ} = 4\epsilon\left(\left(\frac{\sigma}{r}\right)^{12}-\left(\frac{\sigma}{r}\right)^6\right)
\end{equation}
We use a cutoff of $2.5\sigma$.  We simulate the finite temperature and kinetics of the polymer chain dynamics using the Langevin Thermostat
\begin{equation}
m\frac{dV}{dt}=-\Upsilon V-\nabla\Phi(X)+\sqrt{2k_BT\gamma}\frac{dB_t}{dt}.
\end{equation}
The $X$ denotes the position of the atoms, $V= {dX}/{dt}$ is the velocity, $-\nabla\Phi(X)$ denotes the interaction forces, $\Upsilon$ denotes the friction coefficient, and $\sqrt{2k_BT\gamma}dB_t/dt$ denotes the random force accounting for thermal fluctuations~\cite{Gardiner1985}.  We perform all simulations using the LAMMPS molecular dynamics package and our custom extension packages~\cite{Plimpton1995,AtzbergerLAMMPS2016}.

To study the bulk mechanics of the polymeric system, we perform rheological studies using oscillatory shearing motions based on Lees-Edwards boundary conditions~\cite{Lees1972}.  In Lees-Edwards conditions, periodic boundary conditions are used with shifted image interactions.  We use a sinusoidal oscillation of the displacement $L(t)=L_0 + A\sin(2\pi{t}/{T_p})$ with amplitude $A$ and time periodicity $T_p$.  This corresponds to a cosine oscillation of the strain with rate $\dot\gamma=\dot\gamma_0\cos(\omega t)$ where $\omega  = 2\pi/T_p$ and $\dot{\gamma}_0 = A\omega$.

As a measure of material response, we consider the dynamic complex modulus $G(\omega)=G_1(\omega)+iG_2(\omega)$.  The components are defined from measurements of the stress as the least-squares fit of the periodic stress component $\sigma_{xy}$ by the function $g(t)=G_1(\omega)\gamma_0\sin(\omega t)+G_2(\omega)\gamma_0\cos(\omega t)$. This offers a characterization of the response of the material to oscillating applied shear stresses and strains as the frequency $\omega$ is varied. The $G_1$ is referred as the Elastic Storage Modulus and $G_2$ is described as the Viscous Loss Modulus. These dynamic moduli are motivated by considering the linear response of the stress components $\sigma_{xy}$ to applied stresses and strains.
At low frequency the polymer stresses appear to have sufficient time to equilibrate to the applied shear stresses. At high frequencies, the polymer stresses do not appear to have sufficient time to equilibrate to the applied shear stresses. This is manifested in $\sigma_{xy}(t)$ which is seen to track the applied stress very closely. A phase lag 0 is representative of solids and $\pi/2$ is representative of liquids.
This delay is caused by propagating the stress through the domain via the chain topology. The increase of $G_2$ indicates that the mechanics arises effectively from chains' resistance to more rapid motions, such as sliding, while the increase of $G_1$ indicates in the mechanics a resistance to direct deformation represented by increases in the elastic bond lengths or from the bending stiffness of chains.

To estimate the dynamic complex modulus in practice, the least-squares fit is performed for $\sigma_{xy}(t)$ over the entire stochastic trajectory of the simulation after some transient period of approximately $10T$ (see Table \ref{tableparams3}), which is of the order of the diffusion time of the open chains under study. In our simulations, the maximum strain over each period was chosen to be half the periodic unit cell in the $x$-direction, corresponding to a strain amplitude $\gamma_0=\frac{1}{2}$.
A description of the parameters and specific values used in the simulations can be found in Tables \ref{tableparams1} and \ref{tableparams2}. We notice that the applied strain is large and would imply a nonlinear regime for polymer melts. However, the systems considered in this study are polymer solutions of very low molecular weight and our results indicate a linear regime so we can neglect higher harmonic contributions to stress \cite{Vladkov2006}.

The effective stress tensor associated with the polymers at a given time is estimated using the Irving-Kirkwood method \cite{Doi1986,Irving1950}

\begin{equation}
\sigma_{l,k}=\frac{1}{V} \sum_{n}\sum_{j=1}^{n-1}\left\langle f_{j}^{(l)}\cdot(x_{q_n}^{(k)}-x_{q_j}^{(k)})\right\rangle
\end{equation}

\noindent where $V$ is the volume of the periodic box, $x_{q_v}^{(k)}$ is the $k$-th coordinate of the $q_v$-th atom (the minimum image convention is used for the difference) and $f_{j}^{(l)}$ is the $l$-th coordinate of the pairwise interaction between the two atoms.

\begin{table}[H]
\centering
\begin{tabular}{|l|l|l|}
\hline
\textbf{Parameter}  & \textbf{Description}  & \textbf{Value} \\
\hline
$\sigma$   & monomer radius          & 1.0 nm \\
$\epsilon$ & energy scale            & 2.5 $amu\cdot nm^2/ps^2$\\
$m_0$      & reference mass          & 1 amu \\
$w_c$      & energy potential width  & $2.5 \sigma$ \\
$m$        & monomer mass            & 240 $m_0$ \\
$\tau$     & LJ-time-scale           & $\sigma\sqrt{m_0/\epsilon}$ = 0.6 ps\\
$k_B{T}$   & thermal energy          & 1.0 $\epsilon$\\
$\rho$     & solvent mass density    & 39 $m_0/\sigma^3$ \\
$\mu$      & solvent viscosity       & 25 $m_0/\tau\sigma$ \\
$\Upsilon$ & drag coefficient        & 476 $m_0/\tau$  \\
\hline
\end{tabular}
\caption{Parameterization for the polymer weave models.}
\label{tableparams1}
\end{table}

\begin{table}[H]
\centering
\begin{tabular}{|l|l|l|}
\hline
\textbf{Parameter}  & \textbf{Description}  & \textbf{Value} \\
\hline
$E_b$             & harmonic bonds potential constant   & 619.5 $amu/ps^2$ \\
$b$               & harmonic bonds rest length          & 1.0 $nm$ \\
$E_{\theta}$      & harmonic angle potential constant   & 19.8 $amu\cdot nm^2/ps^2$ \\
$\theta_0$        & harmonic angles' rest length         & $180^o$ \\
\hline
\end{tabular}
\caption{Parameterization for the stiffness and connectivity of the polymer chains.}
\label{tableparams2}
\end{table}

Table \ref{tableparams3} shows how the simulation time and oscillation period range compare to characteristic times in our systems (computed using the parameters used in our simulations, shown in the previous tables). The advection time is computed as $\tau_A=m/\Upsilon$. The Rouse time, $\tau_R$, is computed for an ideal chain of length 20. The critical time $\tau_0$, refers to the characteristic time where cross-overs are observed in our simulations, shown in Section \ref{res}. Notice that we do not report an entanglement time because our chains are short (with number of discrete local topological constraints $Z<2$ in many cases) and the notion of entanglement length does not apply to them.

\begin{table}[H]
\centering
\begin{tabular}{|l|l|l|l|l|}
\hline
\textbf{Parameter}              & \textbf{Description}                                & \textbf{value}\\
\hline
$\tau_A$ advection time         & propagation in fluid                                & 0.0013 ps           \\
$\tau_D$ diffusion time         & monomer moves a dist. $\sigma$                      & 302 ps           \\
$\tau_{R}$ Rouse time           & ideal chain $N=20$      & 6937 ps        \\
$\tau_0$ critical time        & cross-over reference time & 598 ps\\
$T$                             & period of oscillation                               & $6ps<T<3600ps$    \\
$t$ simulation time             & longest simulation time                    & $150ps<t<90000ps$ \\
\hline
\end{tabular}
\caption{Characteristic time scales}
\label{tableparams3}
\end{table}

\section{Bulk Mechanical Responses}\label{res}

\subsection{Complex modulus}

\begin{figure}[H]
\centering
\includegraphics[width=0.9\columnwidth]{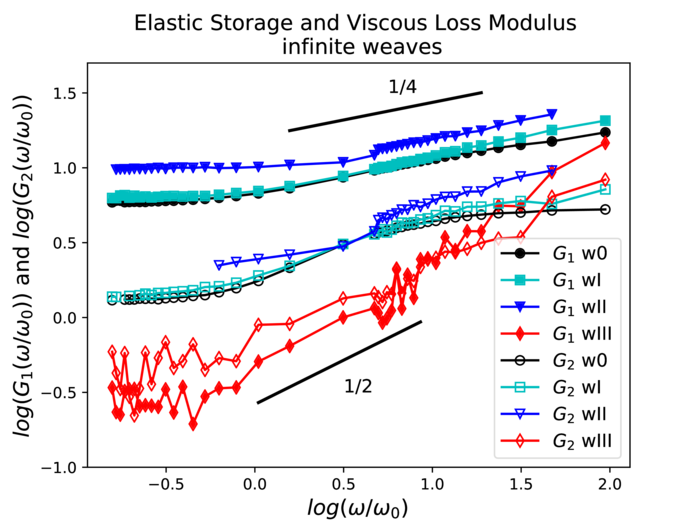}\\
\includegraphics[width=0.9\columnwidth]{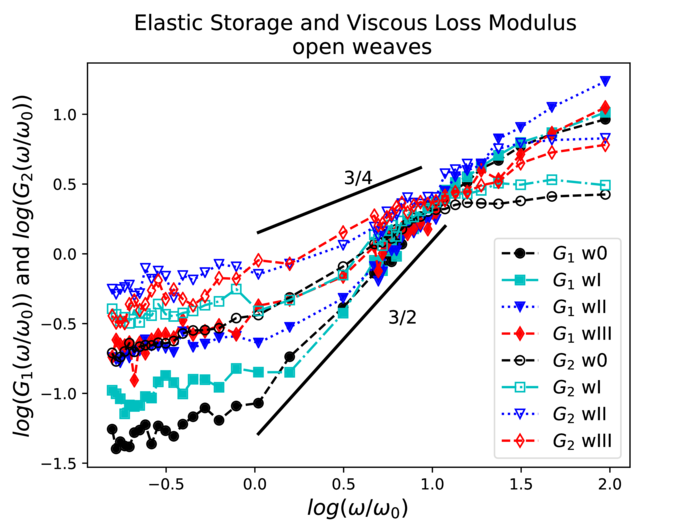}\caption{Polymer Weave Frequency Response: Dynamic Moduli. The Elastic Storage and Viscous Loss Moduli of the infinite chain weaves are shown above and those of the open chain weaves below. The infinite weaves behave like crosslinked polymers with a primarily elastic behavior throughout the range of frequencies simulated. The exponents 1/2 and 1/4 are similar to those in the Rouse model\cite{Edwards2000}. The open weaves transition from an elastic to a viscous behavior as frequency is varied. The exponents 3/4 and 3/2 indicate the predicted scaling for semi-dilute solutions of semi-flexible chains and for the BEL model respectively \cite{Edwards2000}. The slope increases with decreasing topological complexity.}
\label{fig:modulioc}	
\end{figure}

We show the log-log plot of the Elastic Storage Modulus $G_1$ and Viscous Loss Modulus $G_2$ for all the infinite and open weave polymeric materials as the shear response frequency is varied in Figure \ref{fig:modulioc}.  The frequency of oscillation is normalized by $\omega_0=2\pi/\tau_0$ where $\tau_0= 943\tau= 598 ps = 1.98\tau_D$ is a time-scale on the order of the diffusion time $\tau_D$ (see Table \ref{tableparams3}).

Comparing $G_1, G_2$ for the infinite weaves we see that, in the range of frequencies studied, we have $G_1>G_2$ for all the simple weaves.
The crossover of $G_1$ and $G_2$ is absent for those systems within this range of frequencies, which indicates no behavioral transition in the samples which exhibit solid properties.  When $G_1$ is larger than $G_2$ the elastic response is dominant indicating there is relatively few polymer rearrangements (reptation) within the network structure. This indicates that energy is mainly stored elastically in the stretching and bending of bridging polymeric chains. This can be verified by our \textit{Writhe} quantity for the chains as it reaches a minimum at the extrema of the oscillatory strain period within this regime (see Section \ref{confan}).
The systems with large $G_1$ behave like stiff materials having strong entanglements similar to imperfect networks having transient covalent crosslinking \cite{Rosa1994,Hyun2002,Trappe2000,Liew2014,Murphy1986,Clark1987,Mortazavi2003}.
This indicates that polymer solutions of long linear semiflexible chains can behave like crosslinked networks, even in the absence of explicit crosslinks.

Initially, $G_1$ and $G_2$ are independent of the frequency of oscillation and we see a crossover at frequency $\omega_0$ that corresponds to period $\tau_0$. At frequencies higher than $\omega_0$ (period times shorter than $\tau_0$) there is a significant dependence of moduli on the frequency which increases with increasing topological complexity.  This is in agreement with predictions for polymeric networks \cite{Edwards2000}. The line segments shown in the figure indicate a scaling between $\sim\omega^{1/4}$ and $\sim\omega^{1/2}$, respectively, to be compared with that of Rouse chains.

The alternate interlacing weave, wIII, is the only infinite weave for which $G_1$, $G_2$ intersect and for which $G_1$ and $G_2$ both seem to scale as $\omega^{1/2}$ in the intermediate frequencies.  Moreover, for wIII, $G_1\approx G_2$, with $G_1<G_2$ for low frequencies. We find that the original configuration of wIII is not favorable to the stiffness of the chains and the chains need to stretch resulting in a larger $G_1$. This causes extra collisions with other chains which results in larger values of $G_2$ as well. At high frequencies, we notice a shift from
filament bending to stretching which results in higher values of $G_1$. Such transitions have also been observed in networks of actin filaments \cite{Lieleg2009,KimT2009}.

Comparing $G_1,G_2$ for the open systems we find that both $G_1$ and $G_2$ are initially constant up to $\omega\approx\omega_0$ and then increase and intersect at $\omega\approx10\omega_0$. We have $G_2<G_1$ for $\omega<10\omega_0$ and $G_1<G_2$ for $\omega>10\omega_0$. This suggests two critical times in the polymer chain dynamics.  The first is $\tau_e=\tau_0/10$ and the second is $\tau_0$ at which we find have a trend of slightly increasing with decreasing density of the systems as predicted in \cite{Mason1998}.  We find that with increasing frequency the response tends to become dissipative.

At low frequencies $G_1\sim\omega^{1/2}$ we find the trends follow the Rouse model. For larger frequencies we find that $G_1\sim\omega^{3/2}$, $G_2\sim\omega^{3/4}$ and then $G_1, G_2$ tend to a plateau value. Similar scalings were reported in \cite{Wilson2015} and in \cite{Vladkov2006} for polymer solutions of linear FENE chains of similar molecular weight, which further shows that the use of harmonic bonds does not significantly  influence our findings. We find a decreasing slope of $G_1$ for increasing entanglement which suggests a slower relaxation mechanism.

\begin{figure}[H]
\centering
\includegraphics[width=1.0\columnwidth]{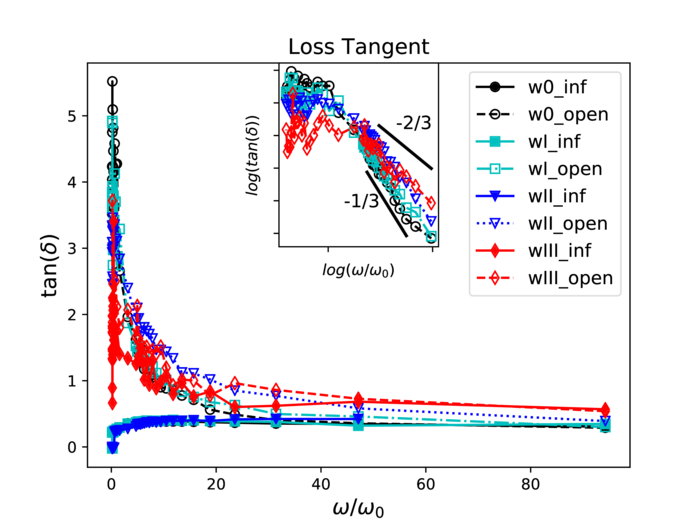}\caption{Polymer Weave Frequency Response: Loss Tangent. The infinite systems behave like crosslinked polymers with a loss tangent less than 1 at all frequencies. The open chains transition from a liquid-like behavior to that of a solid-like behavior as the frequency increases. The inset plot shows the log-log plot for open chains. These results show that the crossover frequency increases with decreasing topological complexity. Similarly, the slope of decrease increases with decreasing topological complexity.}
\label{fig:losstan}	
\end{figure}

We show the loss tangent as a function of the frequency of oscillation in Figure \ref{fig:losstan}. We remark that $\tan\delta$ can be interpreted as reflecting the strength of what is sometimes called ``colloidal forces''.  In other words, if $\tan\delta<1$ then the particles are highly associated and sedimentation could occur. If $\tan\delta>1$, the particles are highly unassociated. The loss tangent is almost constant, close to 0, for all the simple infinite weaves (w0,wI,wII).
The values of the open weaves are greater than one and then decrease to the values of the corresponding infinite weaves. The asymptotic ordering of the phase lag of the systems is $w0<wI<wII<wIII$.

We find that, our data at larger frequencies that all the materials behave like elastic solids, as is often seen in large frequency responses.
The inset graph shows the corresponding log-log plot only for the open systems. It reveals a cross-over at approximately $\omega_0$, which corresponds to times on the order $\tau_0$. This time-scale could be related to the entanglement time as in \cite{Edwards2011,Szanto2017}. This characteristic time-scale, seems to decrease with the topological complexity of the weave.
The large frequency tail of $\tan\delta$ decreases more slowly with increasing topological complexity and density indicating a substantial dissipation effect related to entanglement.

\subsection{Conformational analysis}\label{confan}

\begin{figure}[H]
\centering
\includegraphics[width=1.0\columnwidth]{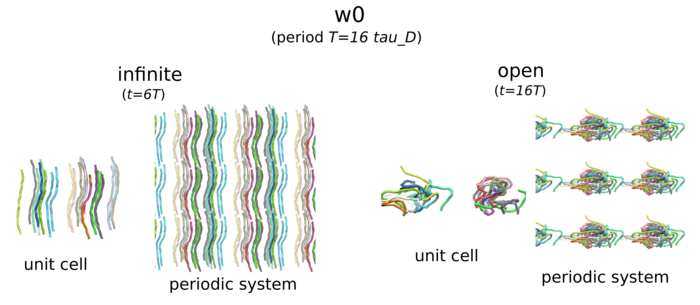}\\
\includegraphics[width=1.0\columnwidth]{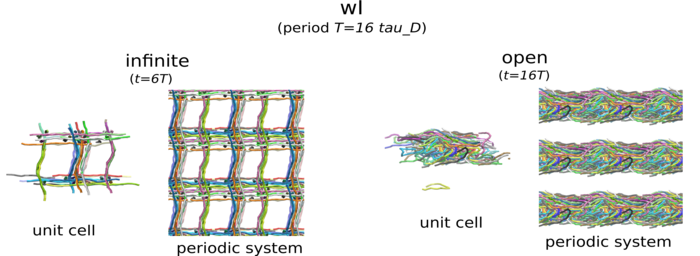}\\
\includegraphics[width=1.0\columnwidth]{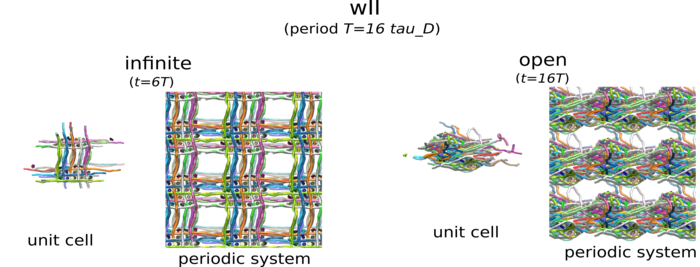}\\
\includegraphics[width=1.0\columnwidth]{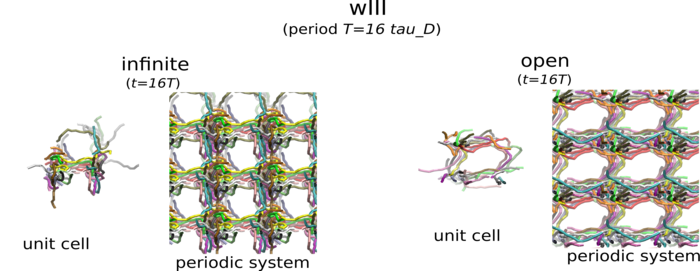}
\caption{Polymer configurations subject to oscillatory shear. Configurations at the end of simulations for the closed chain and open chain systems. In both cases, the chains tend to form bundles. By forming tubes of larger radius the chains decrease their individual deformation. For weave0, the open chains can significantly rearrange their conformations to those of random coils and the system becomes inhomogeneous, disconnected accross the periodic boundary. As the entanglement complexity and density increases, the open chains cannot fully escape their original conformations, forming tubular connected domains (weaveI), lamellar structures (weaveI) and even retain entanglement percolation in three dimensions (weaveIII).}
\label{fig:lastconfig0}
\end{figure}

We show configurations of the polymer weaves at different times during deformation in Figure \ref{fig:lastconfig0}.
For the small oscillation frequencies, the infinite chains follow the deformation of the defining box, attaining an s-shape conformation.
The open chains, significantly rearrange in time and tend to avoid the boundary by aligning with the orientation of the deformation. This process happens more slowly for wI and even more slowly for wII and wIII systems due to topological obstacles.
We note that the chains tend to form bundles of chains, giving an inhomogeneous material, suggesting that the inhomogeneity decreases with increasing density and entanglement complexity.  Similar phase separation of polymer solutions in oscillatory shear has been observed experimentally in \cite{Saito2003}. 

We find the transition from bundle-dominated structures to entanglement dominated structures is related to the entanglement of the chains as has been also reported in \cite{Shen2017}. A possibility for the bundle formation is finite-size effects. To examine that, we performed similar simulations in equilibrium in the NVE and in the NPT ensemble for the w0 infinite system. In both cases, bundle formation occured rapidly in the simulation. This indicates that the bundle formation is not a finite size effect. 

We propose that the chains bundle together in order to decrease their deformation which occurs due to thermal fluctuations and due to the deformation of the cell. As the chains bundle, they form tubes of chains. A larger diameter tube resists the deformation stronger than the individual chains. This larger tube structure is apparent in the w0 infinite system. The same happens in two and three directions respectively for wI, wII and wIII. This also happens transiently at initial times for the open systems. At first the open chains form these bundles then they keep rearranging and entangling further until finally forming globules.

Our topological methods can be used to more precisely characterize the conformational features of the chains.  We do this by measuring the \textit{Writhe} and the \textit{Periodic Linking Number} topological quantities that we introduced in Section~\ref{LK}.  We remark that the chains in our system are loosely entangled relative to highly knotted systems yielding, as a consequence, Periodic Linking Numbers and Writhe that are less than one.  While the quantities appear small relative to those of highly knotted systems, they still provide a useful characterization of the collective configurations of the polymeric chains of the material and their rearrangements.  As our results indicate these are useful in understanding the connection between topology and mechanical responses even in systems of low molecular weight (below that of the entanglement length).

We find a very different behavior of the Writhe for the weave of type w0  between the open and infinite systems (the situation is similar for wI and wII).  In the case of the infinite periodic systems, the mean absolute Writhe of the chains shows a sinusoidal behavior.  This is seen most clearly when the frequency of oscillation is small. During the shearing cycle of the unit cell in our simulations, the Writhe is maximum when the shear deformation is the least and the Writhe reaches a minimum when the shear deformation is at its greatest.  This behavior is indicative of the chains stretching at the maximum deformation and relaxing to a more entangled state when the shear deformation is relaxed.

In the case of open systems, the mean absolute Writhe of the chains also follows a sinusoidal behavior, but it changes significantly in time. This happens because the chains are free to attain any possible configuration and tend to disentangle and relax to configurations similar to those of random coils. Indeed, the final values are similar to those of a semiflexible random coil of comperable length as reported in \cite{Panagiotou2013b}. This behavior becomes less pronounced as we increase the density and complexity of the weave because the disentanglement time increases and the chains do not have sufficient time to rearrange.

\subsection{Topology and Mechanical Responses}
We investigate both closed and open chain systems.  The closed chain systems were found to be less interesting since the values of the Writhe and the Periodic Linking Number did not change significantly as a function of the frequency of oscillation. This is as expected, since without breakage of bonds the topology must remain close to that of the original configuration. Similarly, the loss tangent of the infinite systems does not change significantly throughout the experiment.

\begin{figure}[H]
\centering
\includegraphics[width=1.0\columnwidth]{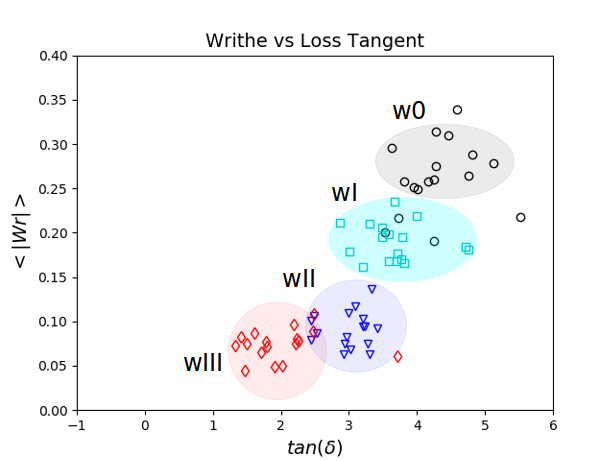}
\caption{Writhe and Loss Tangent of open chains for small frequencies of oscillation (frequencies corresponding to periods $T>6\tau_D$). We find a linear behavior of the Writhe as a function of the Loss Tangent. The data points corresponding to the different weaves form clusters. As the topological complexity of the weave decreases both the Writhe and Loss Tangent increase.}
\label{fig:wrlosstan}	
\end{figure}

Here, we focus on the open chains systems where there is the potential for significant rearrangements of polymer chain configurations and topology.  We show the mean absolute Writhe of open chains as a function of the loss tangent for small frequencies of oscillation in Figure \ref{fig:wrlosstan}.  

We find a decay of the mean absolute Writhe with the loss tangent and a clustering of the data for each system. It is notable that the proposed Writhe measure groups together the systems of similar material response.  The clustering observed for these systems indicates that the global entanglement of the chains imposed by the original conformation affects the response of the material significantly. These results show a relationship between Writhe and $\tan\delta$ that scales like $\langle|Wr|\rangle\sim\tan\delta^{6/5}$.  The responses at small frequencies are clustered and ordered with their Writhe decreasing as $w0>wI>wII>wIII$.  

We remark that as the frequency increases we found the Writhe of the open systems tended to meet the values of their corresponding closed (infinite) chain versions. This occurs since the chains cannot escape their original configuration as readily at large oscillation frequencies.  For the $wIII$ case, approximately the same values for open and closed chains were found throughout.

Previous studies have shown that there is an almost linear relationship between the number of kinks per chain and the mean absolute Writhe of a chain in the case of a melt of linear FENE chains~\cite{Panagiotou2013b}.  The relationship emerges as a function of the molecular weight in equilibrium conditions.  

\begin{figure}[H]
\centering
\includegraphics[width=0.9\columnwidth]{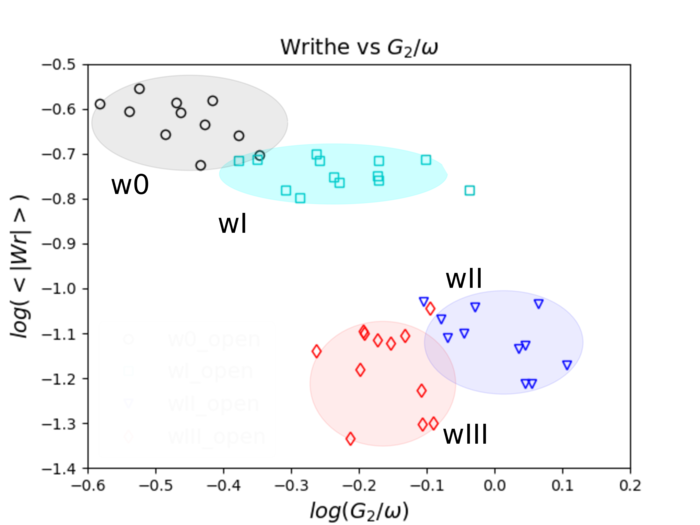}\\
\includegraphics[width=0.9\columnwidth]{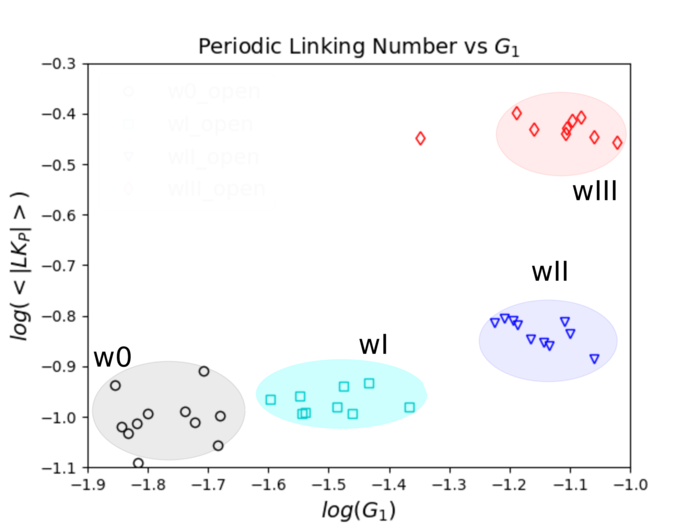}
\caption{Above: Writhe and $G_2/\omega$ for open weaves at small frequencies (log-log plot). The viscosity can be obtained as the limit $\eta=\lim_{\omega\rightarrow0}{G_2(\omega)}/{\omega}$. We find an indication that the Writhe decreases with viscosity, while $Z$ shows the opposite relation to viscosity \cite{Baig2010}. Below: Periodic Linking Number as a function of $G_1$ for open chains for small frequencies (log-log plot). The equilibrium modulus can be obtained as the limit $G_{eq}=\lim_{\omega\rightarrow0}G_1(\omega)$. We find an indication that $G_{eq}$ increases with increasing linking, similar to what was reported for rings in \cite{Everaers1996}. }
\label{fig:etaGeq}	
\end{figure}

In non-equilibrium conditions, the relation becomes more complex \cite{Panagiotou2014}.  The viscosity can be obtained as the limit $\eta=\lim_{\omega\rightarrow0}{G_2(\omega)}/{\omega}$. Because the frequencies studied here are relatively large, however, we can examine how the ratio $G_2/\omega$ depends on topology as a reference. Our results for the smallest frequencies indicate that the Writhe of the open chains decreases with $G_2/\omega$, suggesting a decrease of the Writhe with viscosity, see Figure \ref{fig:etaGeq}. The number of kinks has been shown to increase with viscosity \cite{Kim2008,Baig2010}. A similar inverse behavior between the Writhe of the chains and the number of kinks was observed in the initial time of an elongation of the chains \cite{Panagiotou2014}.

\begin{figure}[H]
\centering
\includegraphics[width=1.0\columnwidth]{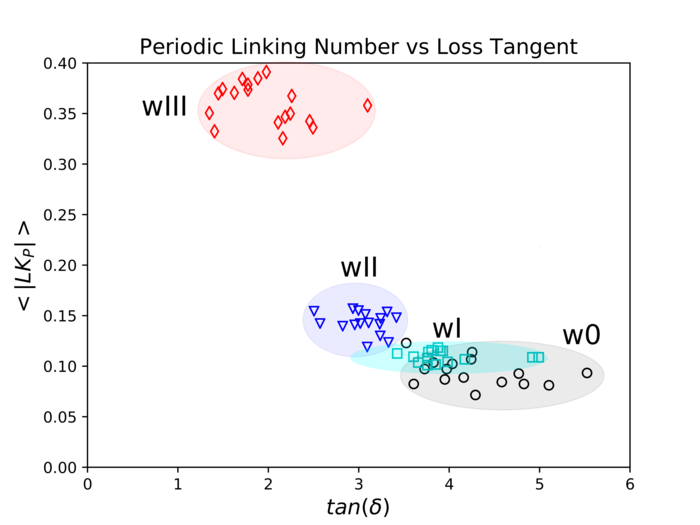}
\caption{Periodic Linking Number and Loss Tangent of open systems for small frequencies of oscillation (corresponding to periods $T>6\tau_D$). We find that the responses corresponding to the different weaves form clusters. The Periodic Linking Number increases with weave complexity and decreases with loss tangent. }
\label{fig:lkplosstan}	
\end{figure}

We show the mean absolute Periodic Linking Number of the open systems at small frequencies as a function of the loss tangent in Figure \ref{fig:lkplosstan}.  We find that the mean absolute Periodic Linking Number decreases with the loss tangent for $\tan\delta>1$. We find significant clustering of the data corresponding to the different systems. The clustering of the responses indicates that the imposed global topology of the initial configuration significantly affects the response of the material.  We remark that with increasing frequency we found the open systems tended to the values of the corresponding infinite systems. This occurs because, at large frequencies, the open chains cannot escape their original configurations.

We see for $\tan\delta>1$ the responses of the open systems can be fit to the form $\langle|LK_P|\rangle\sim\tan\delta^{-5/4}$. The increase of the linking number implies the presence of persistent entanglements.  Persistent entanglements are tight contacts between chains that significantly restrict their motion \cite{Anogiannakis2012}. Such contacts are likely to cause significant bond stretching under deformation that is followed by a decrease of the Loss Tangent. These results indicate that interactions underlying mechanical responses can be effectively captured by the Periodic Linking Number.

From our results we obtain information about the equilibrium modulus. The equilibrium modulus is defined as the limit $G_{eq}=\lim_{\omega\rightarrow0}G_1(\omega)$. We show the storage modulus for the smallest frequencies in Figure \ref{fig:etaGeq}.  Our results suggest that the equilibrium modulus increases with the linking of the chains. Interestingly, a similar linear relation between the entanglement density as obtained from the Gauss linking integral and the shear modulus for ring polymers was reported in \cite{Everaers1996}.

Our results on the Writhe and Periodic Linking Number show a competing relation between these two with respect to the Loss Tangent. We provide a brief explanation for this effect. If the Writhe is large and the Periodic Linking Number is small, this can be interpreted as meaning the chains attain random conformations with no significant topological constraints.  This would result in a behavior that is primarily dissipative.  This suggests that inter-chain contributions to stress dominate through collisions of molecules induced by the Brownian motion.

\begin{figure}[H]
\centering
\includegraphics[width=1.0\columnwidth]{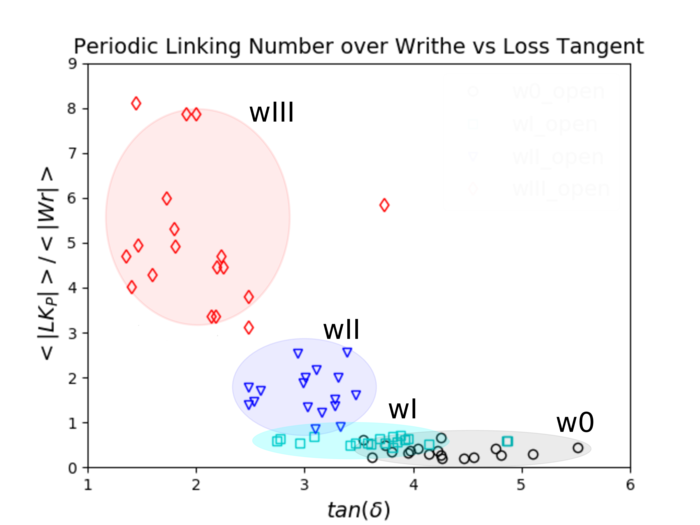}
\caption{The ratio of the Periodic Linking Number over the Writhe as a function of the loss tangent for open systems at low frequencies. We see that the ratio $\langle|LK_P|\rangle/\langle|Wr|\rangle$ decreases with the loss tangent. We also find the data can be fit to a relation of the form  $\langle|LK_P|\rangle/\langle|Wr|\rangle\sim\left(\tan\delta\right)^{-2.6}$.}
\label{fig:lkpwrlosstan}	
\end{figure}

In contrast when the Writhe is small and the Periodic Linking Number is large, we expect that the chains get stretched by the presence of persistent entanglements. In this case, intra-chain contributions to stress would dominate.  As the ratio $\langle|LK_P|\rangle/\langle|Wr|\rangle$ increases, the persistence of entanglements increases. This implies that the bond stretching increases, which decreases the Loss Tangent. Therefore, we expect the loss tangent to increase with the decreasing ratio of Periodic Linking Number versus Writhe.  In fact, this is confirmed in our results as seen in Figure \ref{fig:lkpwrlosstan}.  

Our results show, for the open systems at small frequencies, such a trend of the ratio of the Periodic Linking Number over the Writhe as a function of the Loss Tangent. This indicates that one can control the viscoelastic properties of a material by controlling the ratio of the Writhe and Periodic Linking Numbers of the constituent chains. This also suggests that $\langle|LK_P|\rangle/\langle|Wr|\rangle$ is a measure of the inter-chain contribution versus the intra-chain contribution to the stress.  This finding may contribute to our understanding of the interplay between these two contributions \cite{Sussman2012}.

\section{Conclusions}
We have introduced topological measures for polymeric materials allowing for quantifying the relationship between polymer chain entanglement and viscoelastic responses.  We demonstrated how our topological approaches can be utilized in practice by performing non-equilibrium molecular dynamics simulations of a few polymeric systems.  We found that our topological approaches provide a measure of the molecular-level polymer chain entanglements that contributes to aggregate mechanical responses in the storage modulus and loss modulus.  Our results indicate some of the ways topological tools can be used for characterization of the interplay between entanglement and mechanics.  We expect our methods could be useful in further studies of polymeric materials, such as polymer melts or polydisperse systems with solutions of ring and/or linear chains.

\section{Acknowledgments}
The author P.J.A acknowledges support from research grant NSF CAREER DMS-0956210,
NSF DMS - 1616353, and DOE ASCR CM4 DE-SC0009254.  We also acknowledge UCSB Center for Scientific Computing NSF MRSEC (DMR-1121053) and UCSB MRL NSF CNS-0960316.

\bibliographystyle{plain}
\bibliography{paperDatabase}

\end{document}